\documentclass[twocolumn,showpacs,preprintnumbers,amsmath,amssymb]{revtex4}

\usepackage{amsmath}
\usepackage{natbib}
\usepackage{graphicx}
\usepackage{epstopdf}
\usepackage{subfigure}
\usepackage{dcolumn}
\usepackage{bm}
\usepackage{color}
\begin{document}

\title{\large Chaos of charged particles in quadrupole
magnetic fields under Schwarzschild backgrounds}

\author{Qihan Zhang}
\author{Xin Wu}
\email{21200006@sues.edu.cn; wuxin_1134@sina.com}
 \affiliation{School of Mathematics, Physics and Statistics, Shanghai
University of Engineering Science, Shanghai 201620, China }

\begin{abstract}

A four-vector potential of an external test electromagnetic field
in a Schwarzschild background is described in terms of a
combination of dipole and quadrupole magnetic fields. This
combination  is an interior solution of the source-free Maxwell
equations. Such external test magnetic fields
cause the dynamics of charged particles around the black hole to
be nonintegrable, and are mainly responsible for chaotic dynamics
of charged particles. In addition to the external magnetic fields,
some circumstances should be required for the onset of chaos. The
effect of the magnetic fields on chaos is shown clearly through
an explicit symplectic integrator and a fast Lyapunov indicator.
The inclusion of the quadrupole magnetic fields easily induces
chaos, compared with that of the dipole magnetic fields. This
result is because the Lorentz forces from the quadrupole magnetic
fields are larger than those from the dipole magnetic fields. In
addition, the Lorentz forces act as attractive forces, which
are helpful to bring the occurrence of chaos in
the nonintegrable case.

\end{abstract}
\maketitle

\section{Introduction}

Many observational evidences have supported the presence of
magnetic fields around astrophysical black holes. For example, a
strong magnetic field exists in the vicinity of the supermassive
black hole at the center of the Galaxy [1]. Recently, the
existence of a highly regular, strong magnetic field in the
vicinity of the central, supermassive black hole of 3C 84 was
shown in Ref. [2]. These magnetic fields are due to the dynamics
of ionized matter and plasma in accretion disks surrounding the
black holes. They might be helpful to transfer the energies from
the accretion discs to relativistic jets.

An exterior asymptotically uniform test magnetic field around a
black hole with mass $M$ does not change the gravitational
background when its strength is much smaller than the value
$10^{19}M_\odot /M$ Gauss [3], where $M_\odot$ is the mass of the
Sun. This means that the metric tensor of the black hole geometry
does not have any modification. In other words, such a relatively
weak test magnetic field leads to a negligible effect on the
motion of neutral particles around the black hole. However, it
would exert a very large influence on the motion of a charged test
particle near the black hole. In fact, the relative Lorenz force
is still large if the ratio of the particle's charge to the
particle's mass is large. The motion of charged test particles
should be integrable in a gravitational field like  one of the
Schwarzschild black hole, the Reissner-Nordstr\"{o}m one, the Kerr
one and the Kerr-Newman one, but is most likely nonintegrable in
combined black hole gravitational fields and electromagnetic
fields. There have been a large variety of papers claiming the
onset of chaotic dynamics of charged particles around some black
holes immersed in external magnetic fields (e.g., [4-11]). The
chaotic dynamics close to the black hole horizons is useful to
provide a mechanism for the transmutational energy interchange,
which causes charged particle acceleration to create relativistic
jets [8].

The external test electromagnetic fields around the black holes
are solutions of the vacuum Maxwell equations. One path for the
obtainment of the solutions is from linear combinations of the
spacetime Killing vectors as vector potentials. This is the
so-called Wald's solutions [12], which require that the black
holes should be stationary, axisymmetric, uncharged and vacuum.
However, linear combinations of all Killing vectors with constant
coefficients, as uniform magnetic fields surrounding charged or
nonvacuum black holes (of modified gravity), would fail to satisfy
the source-less Maxwell field equations. In this case, the Wald's
solutions should be generalized by the coefficients taken as
functions of the coordinates [13]. These generalized vector
potentials are expressed in powers of the magnetic field and the
rotation parameter.

A black hole itself can only have a monopole electromagnetic field
in general. The Wald's solutions are only special solutions of the
vacuum Maxwell equations. In fact, the structures of magnetic
fields from white dwarfs, neutron stars, black hole magnetospheres
or external current loops of charged matter like accretion disks
around magnetic compact stars are described by general multipolar
solutions of the vacuum Maxwell equations in curved spacetime
backgrounds [14-18].  A class of general multipolar solutions are
exterior solutions, which are expressed in terms of an infinite
series on the reciprocal of the radial distance [14-19]. They
vanish  at infinity and are based on magnetic coupling to
accretion discs in relation to dynamo action, jets and
hydromagnetic winds. The multipole moments of the currents
circulating in the discs at larger distances from the horizons of
the holes are described via these exterior solutions. Another
class of general multipolar solutions are interior solutions,
which are expressed as polynomials of the radial distance. They do
not vanish  at infinity and should be considered in finite domains
of the radial distances in the inner regions of the magnetic
fields [14-19]. They are in relation to magnetohydrodynamic models
of jet formation and collimation in active galactic nuclei and
microquasars. They describe the contribution of the currents
circulating at the inner edge of the accretion discs at slightly
larger distances from the horizons of the holes. Either the
exterior solutions or the interior ones are combinations of
different multipole magnetic fields.

The innermost stable circular orbits, quasi-harmonic oscillatory
motions and chaotic orbits of charged particles around (general
relativity or modified gravity) black holes, immersed in the
dipole magnetic field as an interior solution of the vacuum
Maxwell equations,  are widely taken into account in the
literature, e.g., [4-11, 20-26].  Nevertheless, the dynamics of
charged particles near the Schwarzschild black hole with the
quadrupole magnetic field as an interior solution of the vacuum
Maxwell equations has been noticed seldom. In this paper, we
mainly focus on  the motion of charged particles in the combined
gravitational field of Schwarzschild black hole and the interior
solution with a combination of dipole and quadrupole magnetic
fields. In particular, the effect of the quadrupole magnetic field
on the charged particle motion is compared with that of the dipole
magnetic field on the charged particle motion.

The rest of this paper is organized as follows. In Section 2, two
types of electromagnetic four-potentials in the Schwarzschild
black hole background are introduced. In Section 3, we are
interested in the motion  of charged particles in the combined
gravitational and electromagnetic background. Finally, the main
results are summarized in Section 4.

\section{Magnetic fields in Schwarzschild geometries}

When the speed of light  and the constant of gravity are taken as
geometric units $c=G=1$, a Schwarzschild black hole with mass $M$
is described in standard Schwarzschild-Droste coordinates
$x^{\alpha}=(t,r,\theta,\phi)$  by the following line element
\begin{eqnarray}
ds^2 = g_{\alpha\beta}dx^{\alpha}dx^{\beta},
\end{eqnarray}
where four nonzero metric components are
\begin{eqnarray}
g_{tt} &=&\frac{1}{g^{tt}}= -\left(1 - \frac{2M}{r}\right), \nonumber \\
g_{rr} &=& \frac{1}{g^{rr}}=  \left(1 - \frac{2M}{r}\right)^{-1}, \nonumber \\
g_{\theta\theta} &=&\frac{1}{g^{\theta\theta}}=  r^2, \nonumber \\
g_{\phi\phi} &=&\frac{1}{g^{\phi\phi}}=  r^2\sin^2\theta.
\nonumber
\end{eqnarray}
This spacetime is static, axisymmetric and asymptotically flat.

Assume that an electromagnetic field exists in the vicinity of the
black hole is too small to affect the spacetime geometry. This
field should be a solution of the source-free Maxwell equations
[16]:
\begin{eqnarray}
0&=&F_{\alpha\beta;\gamma}+F_{\gamma\alpha;\beta}+F_{\beta\gamma;\alpha},
\\
0 &=& J^\mu =\frac{1}{4\pi} F^{\nu\mu}_{;\nu} \nonumber \\
&=& \frac{1}{4\pi\sqrt{-g}}\frac{\partial}{\partial x^\nu}\left(\sqrt{-g} F^{\nu\mu}_{;\nu}\right) \nonumber \\
&=& \frac{1}{4\pi\sqrt{-g}}\frac{\partial}{\partial
x^\nu}\left(\sqrt{-g}g^{\mu\alpha}g^{\nu\beta}
F_{\alpha\beta}\right).
\end{eqnarray}
Here, $J^\mu$ denotes a current,
$g=det(g_{\alpha\beta})=-r^4\sin^2\theta$, and $F_{\mu\nu} =
A_{\nu,\mu} - A_{\mu,\nu}$ is a tensor of the electromagnetic
field. Note that $A_{\nu,\mu}=\partial A_\nu/\partial x^{\mu}$,
where $A_\mu$ is an electromagnetic four-vector potential. For the
case of axial symmetry, the  potential has only one nonzero
component $A_{\phi}$ as a function of  $(r,\theta)$. Thus, the
electromagnetic tensor exists two nonzero components
\begin{eqnarray}
F_{r\phi} = A_{\phi,r}, ~~~~ F_{\theta\phi} = A_{\phi,\theta}.
\end{eqnarray}

The equation $J^\phi =0$ of Equation (3) is rewritten as
\begin{eqnarray}
&&\left(\sqrt{-g}g^{rr}g^{\phi\phi} F_{r\phi}\right)_{,r}+
\left(\sqrt{-g}g^{\theta\theta}g^{\phi\phi} F_{\theta\phi}\right)_{,\theta} \nonumber \\
&&=4\pi\sqrt{-g}J^\phi =0,
\end{eqnarray}
that is,
\begin{eqnarray}
&&r^2 \left[\left(1 -\frac{2M}{r}\right)
A_{\phi,r}\right]_{,r}+\sin\theta
\left(\frac{A_{\phi,\theta}}{\sin\theta}\right)_{,\theta} \nonumber \\
&&=-4\pi r^4J^\phi\sin^2\theta =0.
\end{eqnarray}
This equation is separable and has a series solution [16]
\begin{eqnarray}
A_{\phi}(r,\theta)=\sum^{\infty}_{l=0}\Re_{l}(r)\Phi_{l}(\cos\theta),
\end{eqnarray}
which satisfies two separable equations
\begin{eqnarray}
\Lambda\Re_{l} &=& r^2 \frac{d}{dr}\left[\left(1
-\frac{2M}{r}\right)
\frac{d\Re_{l}}{dr}\right], \\
\Lambda\Phi_{l} &=& -(1-u^2)\frac{d^2\Phi_{l}}{du^2}.
\end{eqnarray}
Note that $\Lambda=l(l+1)$ and $u=\cos\theta$.

Based on the Legendre polynomials $P_{l}(u)$, the angular
functions $\Phi_{l}$ are determined by $\Phi_{0}(u)=0$ and
$\Phi_{l}(u)=l(P_{l-1}(u)-uP_l(u))$ $(l=1,2,\ldots)$. The radial
functions $\Re_{l}(r)$ are an infinite series of $1/r$ [16]:
\begin{eqnarray}
f_l(r)=\sum^{\infty}_{n=l}c^{(l)}_{n}\left(\frac{2M}{r}\right)^n,
\end{eqnarray}
where $c^{(l)}_{n}$ ($n>l$) are a set of
coefficients, expressed in terms of free parameters $c^{(l)}_{l}$
[18]. It is clear that $f_l(r)\rightarrow 0$ as
$r\rightarrow\infty$. Equation (7) with Equation (10) is a class
of exterior solutions of Equation (6) [19].

The radial functions are also expressed in terms of polynomials
with respect to $r$ as
\begin{eqnarray}
g_{l}(r)=\sum_{n=2}^{l+1} a_{n}^{(l)}\left(\frac{r}{2
M}\right)^{n},
\end{eqnarray}
where the coefficients $a_{n}^{(l)}$ are written in [18] as
\begin{eqnarray}
a_{2+k}^{(l)} = \prod_{m=1}^{k} \left(\frac{m(m+1) -
l(l+1)}{m(m+2)} \right) a_2^{(l)}.
\end{eqnarray}
Note that $k = 1, \dots, l-1$ are considered. Equation (7) with
Equation (11) is a class of interior solutions of Equation (6)
[19]. For $l=1$, Equation (7) with Equation (11) corresponds to
the magnetic field potential
\begin{equation}
A_{\phi 1} =g_{1}(r)\Phi_1(u)= a_2^{(1)} \left(\frac{r}{2M}
\right)^2 \sin^2 \theta.
\end{equation}
When  $a_2^{(1)}=2BM^2$, the potential is
\begin{eqnarray}
A_{\phi1} =\frac{1}{2} Br^2 \sin^2\theta,
\end{eqnarray}
where $B$ represents the strength of magnetic field. The result is
consistent with the Wald potential [12], which arises from the
space-like Killing vector $ \xi^\alpha_{(\phi)}=(0,0,0,1)$ through
the relation $A^\alpha = (B/2)\xi^\alpha_{(\phi)}$. Hereafter, the
potential of Equation (14), as an interior solution of Equation
(6), represents a uniform magnetic field. It is labelled as
Magnetic Field 1 (MF1), and is related to the dipole magnetic
field [16,18,19]. For $l=2$, Equation (7) with Equation (11)
stands for the potential from the sum of the uniform magnetic
field and the quadrupole one:
\begin{eqnarray}
A_{\phi 2} &=& g_{1}(r)\Phi_1(u)+g_{2}(r)\Phi_2(u) \nonumber \\
&=&  a_2^{(1)} \left( \frac{r}{2M} \right)^2 \sin^2 \theta+ 3
a_2^{(2)} \left(\frac{r}{2M} \right)^2  \nonumber \\
&& \cdot\left( 1 - \frac{2r}{3M} \right) \cos \theta \sin^2
\theta.
\end{eqnarray}
The term $g_{2}(r)\Phi_2(u)$ corresponds to a quadrupole magnetic
field as an interior solution of Equation (6). Given
$a_2^{(2)}=a_2^{(1)}=2BM^2$ (this choice is
considered because the potential of Equation (13) in the case of
$a_2^{(1)}=2BM^2$ is the Wald potential), the potential is of the
expression
\begin{eqnarray}
A_{\phi 2}=\frac{1}{2}Br^2 \sin^2 \theta
\left[1+\left(3-\frac{2r}{M}\right)\cos\theta\right].
\end{eqnarray}
Such a combination of the two magnetic fields is also an interior
solution of Equation (6), marked as MF2. It is clear that the
potentials $A_{\phi 1}$ and $A_{\phi 2}$ do not tend to zero as
$r$ gets sufficiently large. They should be limited to a finite
interval of $r$.

Why are the two kinds of  potentials $A_{\phi 1}$ and $A_{\phi 2}$
considered in this paper? There are several reasons. Although the
functions $f_l(r)$ and $g_l(r)$ have been given in [18], one
cannot clearly know what physical meanings these coefficients
$c^{(l)}_{n}$ and $a_{n}^{(l)}$ are. How to choose these
coefficients is almost unclear, either. Only the choice
$a_2^{(1)}=2BM^2$ of Equation (13) has appeared in many
references, such as [5-10, 20-26]. It was also reported in [18]
that circular equatorial orbits always exist in the odd-multipole
magnetic fields given by Equations (7) and (10). However, the
coefficients $c^{(l)}_{n}$ associated with the magnetic field
strength $B$ should be extremely large so that the functions
$f_l(r)$ have no negligible contributions for the radial distance
$r$ larger than the black hole horizon $R_h$. Such extremely
strong magnetic fields would be very likely to lead to changes of
the spacetime geometries. On the other hand, the coefficients
$a_{n}^{(l)}$ associated with the magnetic field strength $B$
should be small enough because $r^{l+1}$ becomes extremely large
for $l\geq3$ and $r\gg R_h$. The role of such extremely small
coefficients $a_{n}^{(l)}$ makes the functions $g_l(r)$ have
somewhat important contributions.

The nonzero components of the magnetic fields described by the two
kinds of potentials are written in [18] as
\begin{eqnarray}
B_r =\frac{F_{\theta\phi}}{r^2\sin\theta}, ~~B_\theta
=-\frac{F_{r\phi}}{\sin\theta}\left(1-\frac{2M}{r}\right).
\end{eqnarray}
For $A_{\phi 1}$, the two nonzero components of the magnetic field
are
\begin{eqnarray}
B_{r1} =B_z\cos\theta, ~~B_{\theta1}= -B_z(r-2M)\sin\theta,
\end{eqnarray}
where $B_z=B$ denotes the constant magnetic field strength along
the positive $z$-axis. The potential $A_{\phi 2}$ corresponds to
the two nonzero components of the magnetic field:
\begin{eqnarray}
B_{r2} &=& B_z \left[\cos\theta+\left(\frac{3}{2}-\frac{r}{M}\right)\left(3\cos^2\theta-1\right)\right], \\
B_{\theta2} &=&
-B_z(r-2M)\sin\theta\left[1+3\left(1-\frac{r}{M}\right)\cos\theta\right].
\end{eqnarray}

Although the weak external magnetic fields exert negligible
influences on the spacetime geometries, they play an important
role in the motion of charged particles near the Schwarzschild
black holes.

\section{Dynamics of charged particles in a
combination of dipole and quadrupole magnetic fields}

The motion of charged particles around the Schwarzschild black
holes in the combination of dipole and quadrupole magnetic fields
is mainly focused on. For comparison, the dynamics of charged
particles in the dipole magnetic field is also considered
together.

\subsection{Dynamical equations and numerical schemes}

The motion equations of a charged test particle with charge $q$
and mass $m_p$ are governed in terms of the Hamiltonian system
\begin{eqnarray}
H &=& \frac{1}{2m_p} g^{\mu\nu} (p_\mu - q A_\mu)(p_\nu - q A_\nu)
\nonumber \\
&=& \frac{1}{2m_p}[g^{tt}p^{2}_{t}+g^{rr}p^{2}_{r}
+g^{\theta\theta}p^{2}_{\theta\theta} \nonumber \\
&& +g^{\phi\phi} (p_\phi - q A_\phi)^2].
\end{eqnarray}
For MF1, $A_\phi=A_{\phi1}$. For MF2, $A_\phi=A_{\phi2}$. Because
this Hamiltonian does not explicitly depend on the coordinates $t$
and $\phi$, two momentum components $p_t$ and $p_\phi$ are
constant. They are expressed as
\begin{eqnarray}
p_t &=& m_pg_{tt}\dot{t}=-E, \\
p_\phi &=& m_pg_{\phi\phi}\dot{\phi}+q A_\phi =L,
\end{eqnarray}
where $E$ and $L$ are the energy and angular momentum of the
particle, respectively. In this case, the Hamiltonian is a system
of two degrees of freedom:
\begin{eqnarray}
H &=& \frac{1}{2m_pr^2\sin^2\theta}(L-qA_{\phi
})^{2}-\frac{1}{2m_p}(1-\frac{2}{r})^{-1} E^{2} \nonumber \\
&& +\frac{1}{2m_p}(1-\frac{2}{r})p^{2}_r
+\frac{p_\theta^2}{2m_pr^2}.
\end{eqnarray}
The rest mass in the timelike spacetime corresponds to a conserved
Hamiltonian quantity of the form
\begin{equation}
H = -\frac{m_p}{2}.
\end{equation}

To simplify the related expressions, we apply dimensionless
operations to the Hamiltonian (24) via a series of scale
transformations: $r \rightarrow rM$, $\tau\rightarrow M\tau$,
$t\rightarrow Mt$, $E \rightarrow m_pE$, $p_r \rightarrow m_pp_r$,
$L \rightarrow m_pML$, $p_{\theta} \rightarrow m_pM p_{\theta}$,
$q \rightarrow m_pq$, $B \rightarrow B/M$ and $H \rightarrow
m_pH$. Note that $\tau$ is the proper time. Hereafter, $\beta=Bq$
is used. In this way, the mass factors $M$ and $m_p$ are absent in
all the above-mentioned expressions, and the Hamiltonian (24)
becomes dimensionless. Of course, the correspondences between the
dimensionless qualities and the practical physical qualities can
also be seen from these scale transformations. For instance, the
dimensionless angular momentum $L$ corresponds to the practical
angular momentum $m_pML$. When the dimensionless strength of
magnetic field is $B$, the practical one should be $B/M$.

Because of the inclusion of the external magnetic fields in the
dimensionless Hamiltonian (24), a fourth constant of motion is no
longer existent. This Hamiltonian is not integrable. Numerical
integration schemes are a good technique to study this Hamiltonian
dynamics.

The dimensionless Hamiltonian can be split into four explicitly
integrable parts for the presence of  the external magnetic
fields, as it can for the absence of  the external magnetic fields
in [27]. Hence, a second-order explicit symplectic integrator
$S_2$ and a fourth-order one $S_4$ [28], which preserve the
symplectic structure of the Hamiltonian, are easily available. See
the paper [27] for more details on the construction of these
explicit symplectic methods in the Schwarzschild spacetime.

Figure 1 tests the numerical performance of the two algorithms by
means of three orbits in the two magnetic fields. Here,
both symplectic algorithms $S_2$ and $S_4$ use a
fixed time step $h=\Delta\tau$, where $h=1$. The parameters are
taken as $E =0.995$ and $L=4.5$. Orbits 1 and 2 have their initial
conditions $p_r = 0$ and $\theta = \pi/2$. In MFF1, $\beta =
1.8\times 10^{-3}$, and the initial separations are $r=50$ for
Orbit 1 and $r = 11$ for Orbit 2. The starting values of
$p_{\theta}>0$ in the two orbits are solved from the theoretical
result $\Delta H=H+1/2=0$. However, $\Delta H\neq0$ from the
numerical viewpoint, as shown in Figure 1(a),(b). The method $S_2$
shows no secular growth in the errors $\Delta H$ for both orbits.
This is attributed to a key characteristic of a symplectic
integrator. The secular growth of the errors is still absent for
the algorithm $S_4$ acting on Orbit 2, but does not appear until
Orbit 1 is integrated for a long enough time by $S_4$. There are
no typical differences in accuracies of $\Delta H$ between the two
orbits for each of the algorithms.  $S_4$ performs about three
orders  of magnitude better in the accuracy than  $S_2$. Figure
1(c),(d) shows that the two algorithms have similar performance
for Orbits 2 and 3 in MFF2, where the magnetic field parameter is
$\beta = 3.1 \times 10^{-6}$ and the initial separation is $r =90$
for Orbit 3.

A notable point is that the error curves have more dramatic
changes in Figure 1(a),(c) than those in Figure 1(b),(d). This
phenomenon implies whether Orbit 1 in MF1 and Orbit 2 in MF2 are
possibly different from Orbit 2 in MF1 and Orbit 3 in MF2 in the
orbital dynamical behavior. In what follows, we shall answer this
question using $S_4$ rather than $S_2$ due to the preference of
$S_4$ over $S_2$ in the accuracy.

\subsection{Chaotic dynamics of charged particles in the two electromagnetic potentials}

Based on the analysis of Poincar\'{e} sections, the dynamical
nature of the orbits in Figure 1 can be determined clearly. Orbit
2 in MF1 of Figure 2(a) and Orbit 3 in MF2 of Figure 2(d) are
regular because the intersection points $(r,p_r)$ of the orbits
intersected with the plane  $\theta =\pi/2$ and $p_\theta > 0$
form continuous smooth closed curves. However, Orbit 1 in MF1 of
Figure 2(a) and Orbit 2 in MF2 of Figure 2(d) exhibit chaotic
dynamics because the intersection points randomly fill
two-dimensional regions in the phase space $r-p_r$.

The dynamical nature is also shown through the largest Lyapunov
exponents (LEs), which measures the diverge or converge rate
between nearby trajectories in phase space. They are defined in
[29] by
\begin{equation}
\lambda = \lim_{\tau \to \infty} \frac{1}{\tau} \ln
\frac{d(\tau)}{d_0},
\end{equation}
where \( d(\tau) \) and \( d_0 \) are the proper distances between
two nearby orbits at the time \(\tau \) and the starting time,
respectively. Positive values of $\lambda$ correspond to the
exponential diverge between nearby trajectories and sensitive
dependence on initial conditions. They indicate chaotic dynamics
of bounded  Orbit 1 in MF1 of Figure 2(b) and Orbit 2 in MF2 of
Figure 2(e). Nevertheless, zero values of the LEs mean the
power-law diverge between nearby trajectories and show the regular
nature of Orbit 2 in MF1 of Figure 2(b) and Orbit 3 in MF2 of
Figure 2(e).

It is worth emphasizing that the true values of the LEs are from
the limit values of $\lambda$ as $\tau\rightarrow\infty$. Although
the integration times have no way to tend to infinity from
practical computations, they are still required to be long enough,
e.g., $\tau=10^{8}$. In this case, it takes more CPU times to
carry out this task. Fast Lyapunov indicators (FLIs) [30] are
quicker to distinguish between regular and chaotic dynamics than
the LEs. They can also be calculated in terms of the distances
between two nearby orbits [29] by
\begin{equation}
FLI = \log_{10} \frac{d(\tau)}{d_0}.
\end{equation}
When  the integration time is equal to $\tau=10^{6}$, the FLIs can
clearly describe the regularity of  Orbit 2 in MF1 of Figure 2(c)
and Orbit 3 in MF2 of Figure 2(f) through the FLIs that increase
algebraically slowly with the time $\log_{10}\tau$. The chaoticity
of bounded Orbit 1 in MF1 of Figure 2(c) and Orbit 2 in MF2 of
Figure 2(f) is also shown by  the FLIs that increase
exponentially with time.

Since the technique of FLIs is faster to find chaos than that of
LEs, it is mainly employed to survey  the effect of a small change
of one or two parameters on a transition from regular dynamics to
chaotic dynamics. The magnetic field parameter \(\beta = 0.00185
\) and the initial separation \( r=50\) are considered. Taking the
angular momentum \( L = 4.5 \), we have the correspondence of the
energy $E$ and the FLI in Figure 3(a), where $E$ is varied from
0.990 to 1.00 in an interval of 0.0001. Each of the FLIs is not
obtained until $\tau=10^{6}$.  All values of the energy $E$ with
\( FLIs < 5 \) admit the regular dynamics, whereas those with \(
FLIs \geq 5 \) indicate the chaotic dynamics. As $E>0.993$
increases, chaos is most likely to happen and is strengthened in
MF1. However, the dependence of the FLI on the angular momentum
$L$ in Figure 3(b) shows that  the possibility for the occurrence
of chaos is large for a small value of $L$ in MF1. These results
are also supported by the dependence of the FLI on the two
parameter space $(E, L)$ in Figure 3(c). On the other hand, strong
chaos is always existent for the considered parameters in MF2 of
Figure 3(d)-(f). This extent of chaos seems to be independent of
the values of the angular momentum $L$ in MF2. However, the
occurrence of chaos becomes difficult when the angular momentum
$L$ increases in MF1. This fact sufficiently shows that the
inclusion of quadrupole magnetic fields dramatically enhances the
extent of chaos of charged particles.

The possibility for chaos would increase in MF1 when both the
magnetic parameter $\beta$ and the energy $E$ increase, as shown
via the FLIs corresponding to the two parameters in Figure 4(a).
However, there is a great strong chaotic region in the two
parameter space $(E, \beta)$ for the case of MF2 in Figure 4(b).
The onset of strong chaos is not typically affected, regardless of
whether the energy $E$ and the magnetic parameter $\beta $ (larger
than a certain value) are large or small. On the other hand, chaos
gets easier in MF1 of Figure 5(a) with the increase of $\beta$ or
the decrease of $L$. Chaos is almost allowed for the parameters
$L$ and $\beta$ obtained in MF2 of Figure 5(b). The chaotic
dynamics do not depend on the values of the magnetic parameter
$\beta $ (larger than a certain value) and the angular momentum
$L$ in MF2. It is worth pointing out that chaos is impossible in
Figures 4 and 5 if the magnetic parameter $\beta$ is too small.
This point is clearly shown from the integrable dynamics in the
case of $\beta=0$.

In a word, the possibility and strength for chaos are larger for
MF2 with the combination of dipole and quadrupole magnetic fields
than those for MF1 with the dipole magnetic fields under some
appropriate circumstances. This result is because the contribution
of the quadrupole magnetic fields to the charged particle motions
is larger than that of the dipole magnetic fields. This point can
be seen clearly from Equation (16), where
$2r-3\gg1$ for $r\gg2$ and $\theta=\pi/2$. Namely, the  quadrupole
term plays a dominant role in Equation (15) or (16), compared with
the dipole one. In other words, the results of Figures 3(d)-(f),
4(b) and 5(b) have no typical differences between the exclusion of
the dipole term and the inclusion of the dipole term in Equation
(15) or (16). In addition, both the Lorentz forces from the dipole
magnetic fields and those from the quadrupole magnetic fields act
as attractive forces. Similarly, the energy $E$ acting on the
Newtonian gravitational potential of the black hole also brings an
attractive force contribution to the particles. These attractive
forces are helpful to induce chaos under appropriate conditions
from a statistical viewpoint. However, the angular momentum $L$
gives rise to a repulsive force contribution. Consequently, its
increase would weaken the degree of chaos.

\section{Summary}
\label{sec1}

The four-vector potentials of electromagnetic fields in  the
Schwarzschild background, as solutions of the source-free Maxwell
equations, have two kinds of expressional forms. One kind of
expression is exterior solutions expressed as an infinite series
on the reciprocal of the radial distance, and another kind of
expression is interior solutions expressed as polynomials of the
radial distance. For convenience of practical computations, the
latter kind of four-vector potential with a combination of dipole
and quadrupole magnetic fields in a finite interval of the radial
distance is considered as an interior solution of the source-free
Maxwell equations in this paper.

The motion of charged particles in the combination of dipole and
quadrupole magnetic fields is nonintegrable. Explicit symplectic
integrators that conserve the phase flow of the Hamiltonian system
are very suitable for studying the long-term evolution of the
charged particle motions. One of the integrators is applied to
investigate the effects of the magnetic fields on chaotic dynamics
of charged particles with the fast Lyapunov indicator. Compared
with the inclusion of the  dipole magnetic fields, that of the
quadrupole magnetic fields easily induces chaos under some
circumstances. This is because the external test
magnetic fields cause the dynamics of charged particles to be
nonintegrable and are mainly responsible for chaos of charged
particles. In addition, the quadrupole magnetic fields give a more
contribution to the charged particle motions than the dipole
magnetic fields. In fact, the Lorentz forces from the quadrupole
magnetic fields as well as those from  the dipole magnetic fields
act as attractive forces, which are helpful to encourage the
occurrence of chaos in the nonintegrable case.


\textbf{Author Contributions}: H.Z. made contributions to the
software, and writing-original draft. X.W. contributed to the
supervision, conceptualization, writing-review and editing, and
funding acquisition. All authors have read and agreed to the
published version of the manuscript.

\textbf{Funding}: This research was supported by the National
Natural Science Foundation of China (Grant No. 11973020).

\textbf{Data Availability Statement}: All of the data are shown as
the figures and formula. No other associated data.

\textbf{Conflicts of Interest}: The authors declare no conflict of
interest.

\begin{figure*}[htpb]
\centering{
\includegraphics[width=16pc]{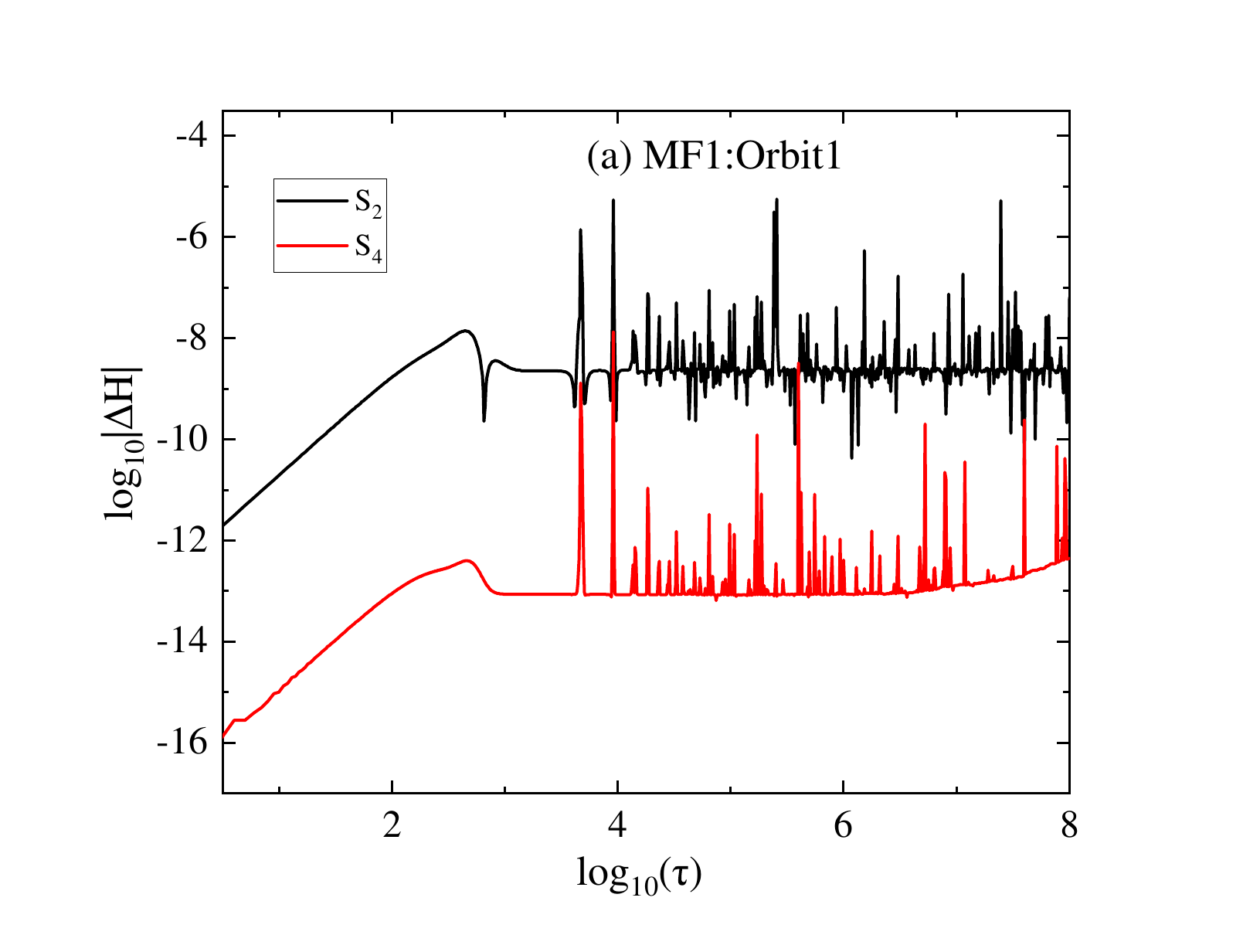}
\includegraphics[width=16pc]{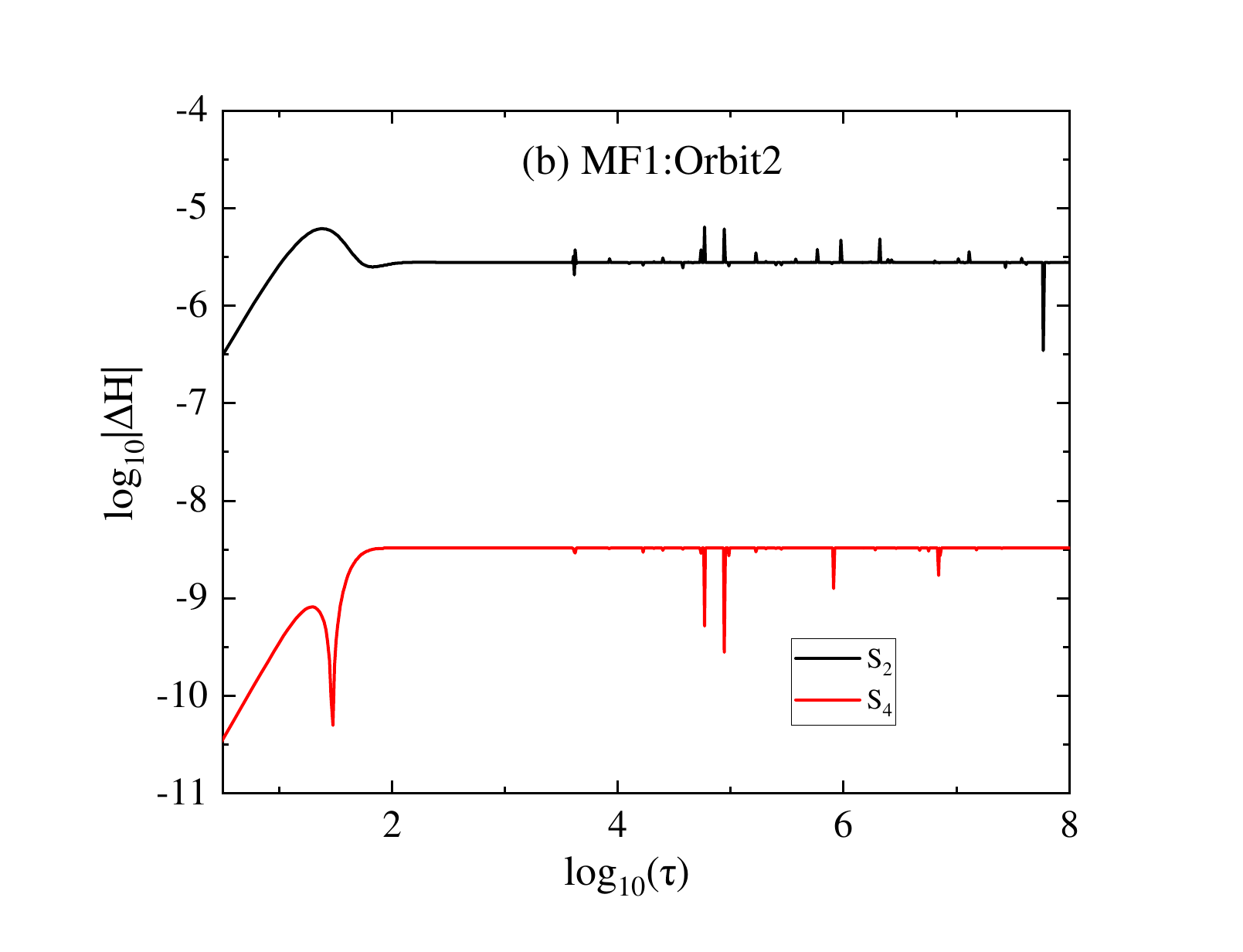}
\includegraphics[width=16pc]{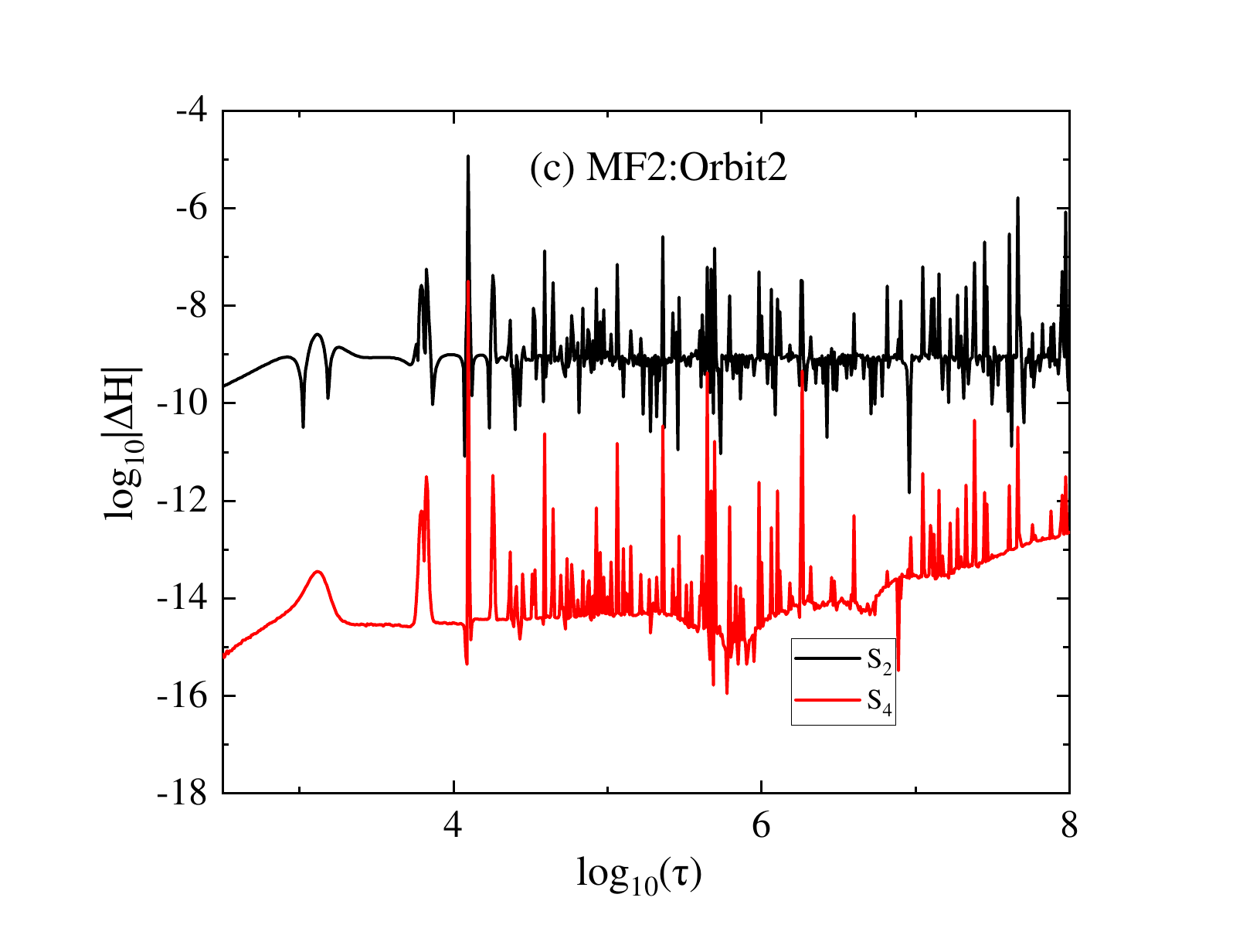}
\includegraphics[width=16pc]{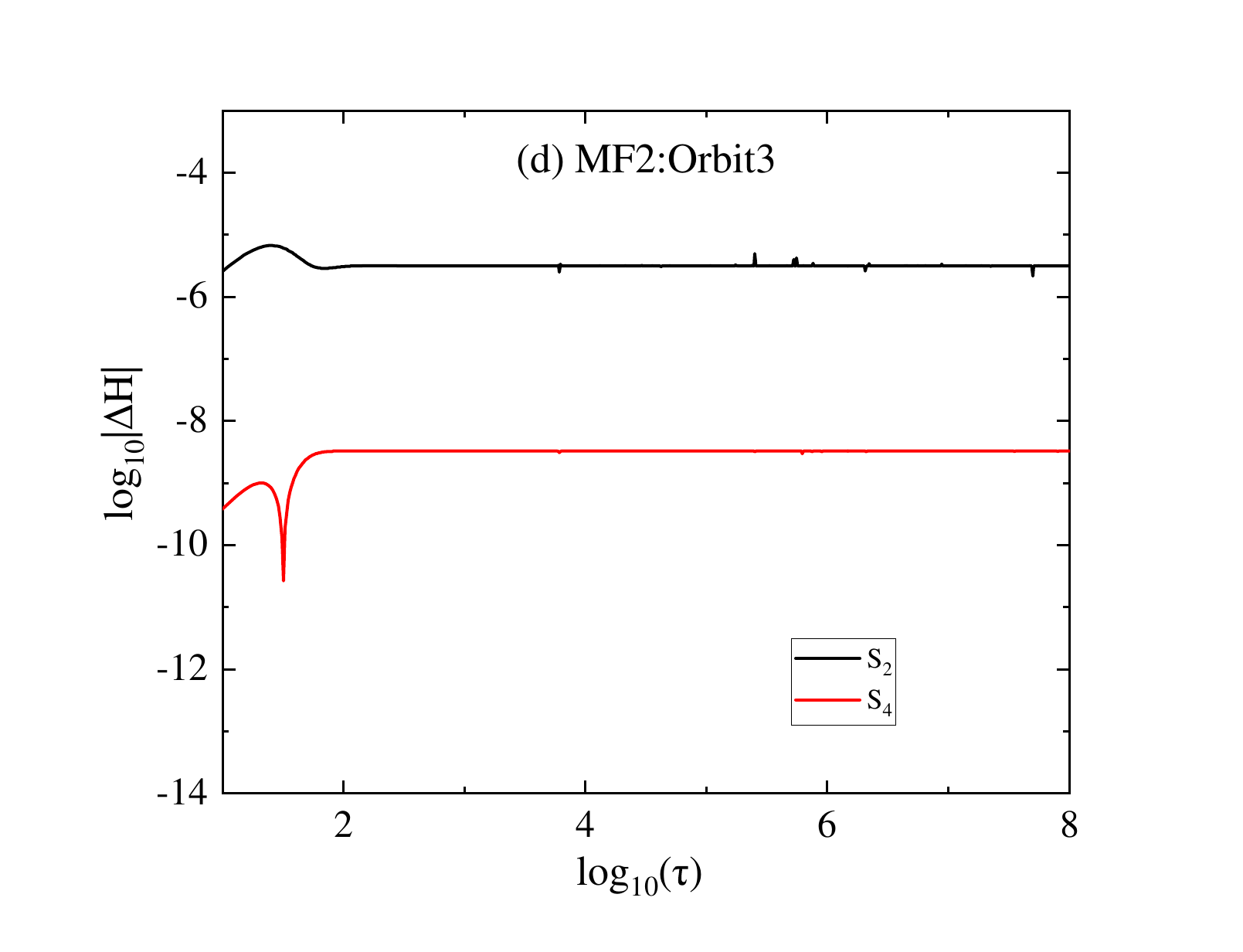}
\caption{Evolution of Hamiltonian errors $\Delta H$ with the
proper time $\tau$ for the methods $S_2$ and $S_4$ in the two
magnetic fields, labelled as MF1 and MF2. The parameters are $E =
0.995$ and $L = 4.5$, and the initial conditions are $p_r = 0$ and
$\theta = \pi/2$. The initial separations are (a) $r = 50$ for
Orbit 1 and (b) $r =11$ for Orbit 2 in MF1 with $\beta =1.8\times
10^{-3}$. The initial separations are (c) $r = 11$ for Orbit 2 and
(d) $r =90$ for Orbit 3 in MF2 with $\beta =3.1\times 10^{-6}$. }
}
\end{figure*}

\begin{figure*}[htpb]
\centering{
\includegraphics[width=13pc]{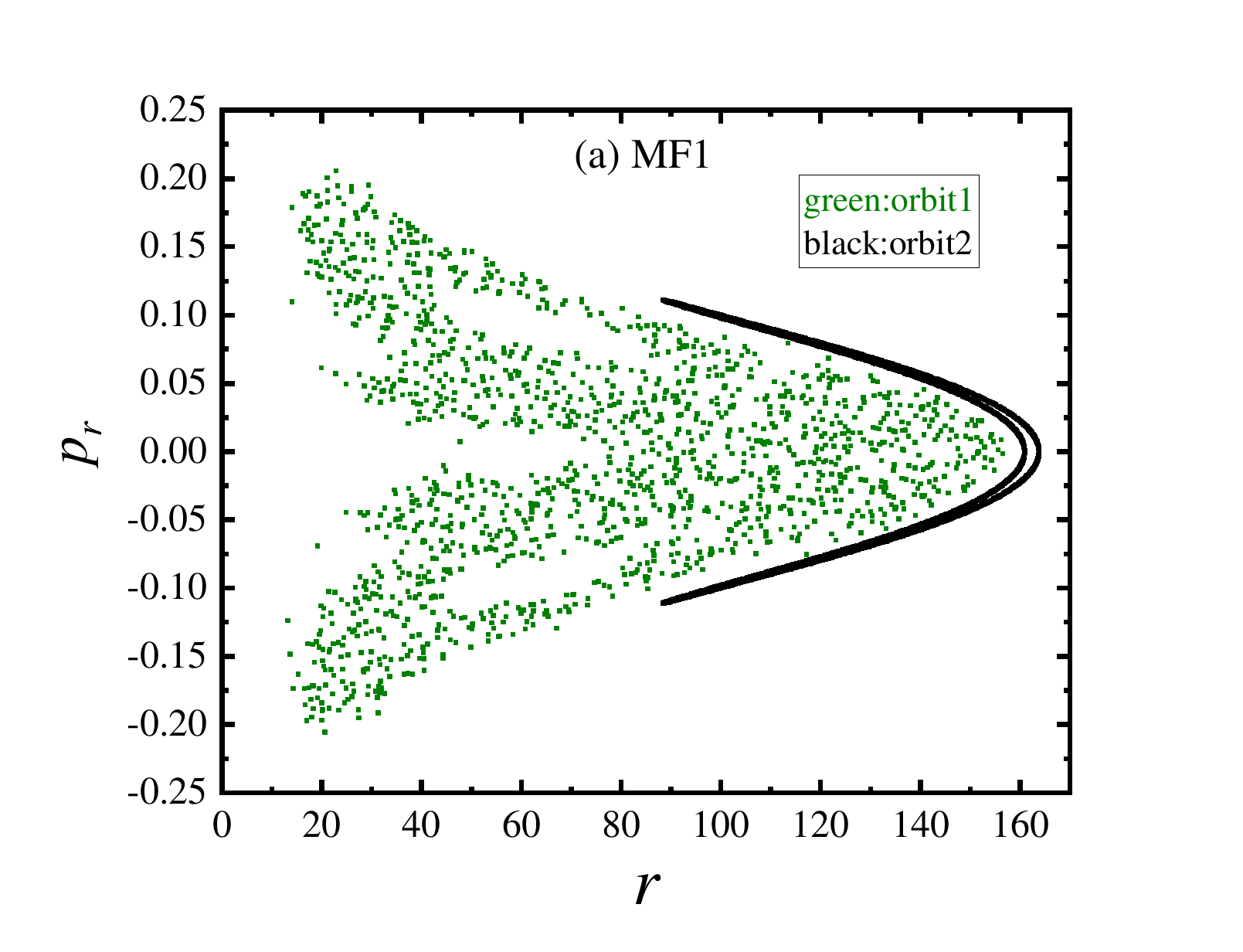}
\includegraphics[width=13pc]{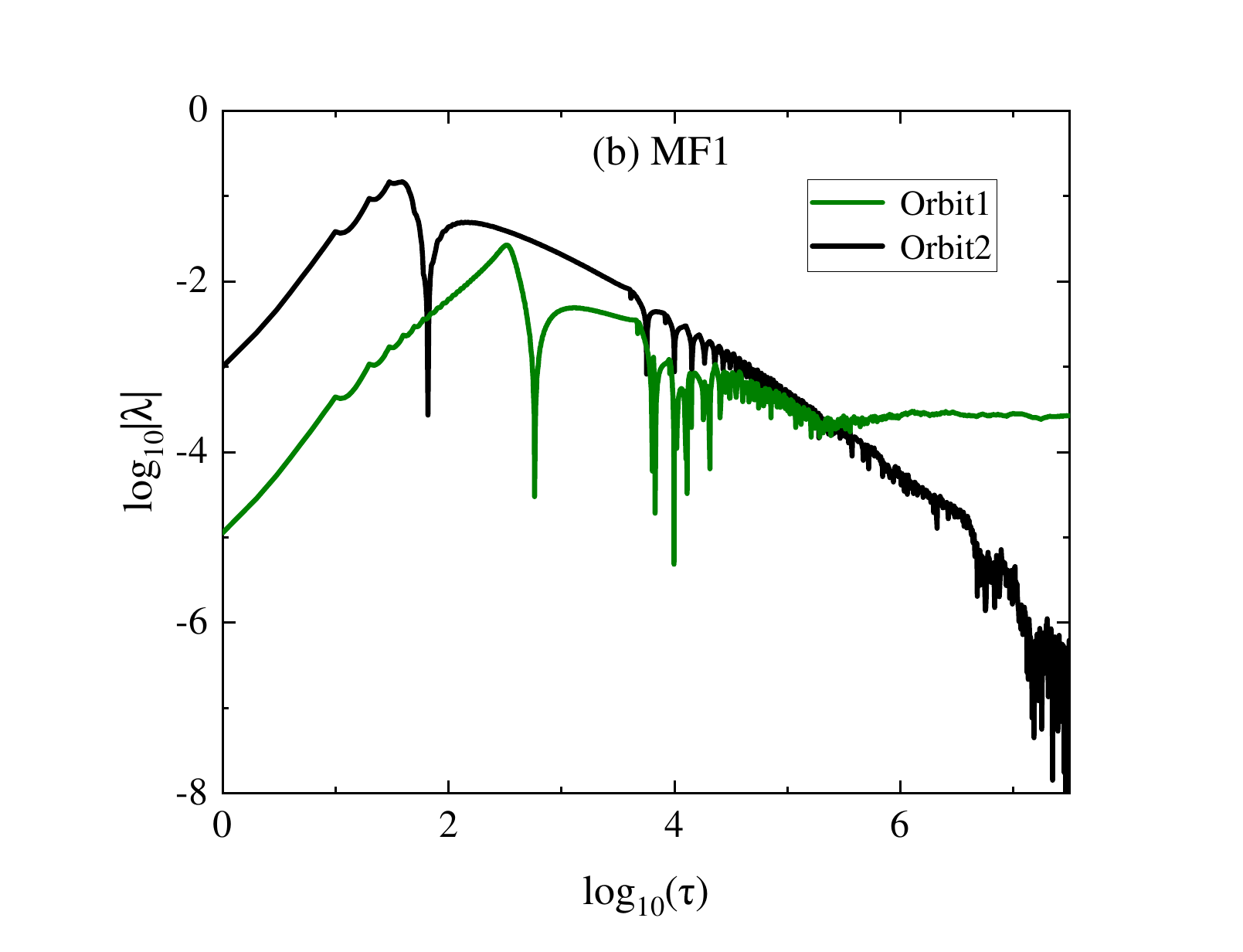}
\includegraphics[width=13pc]{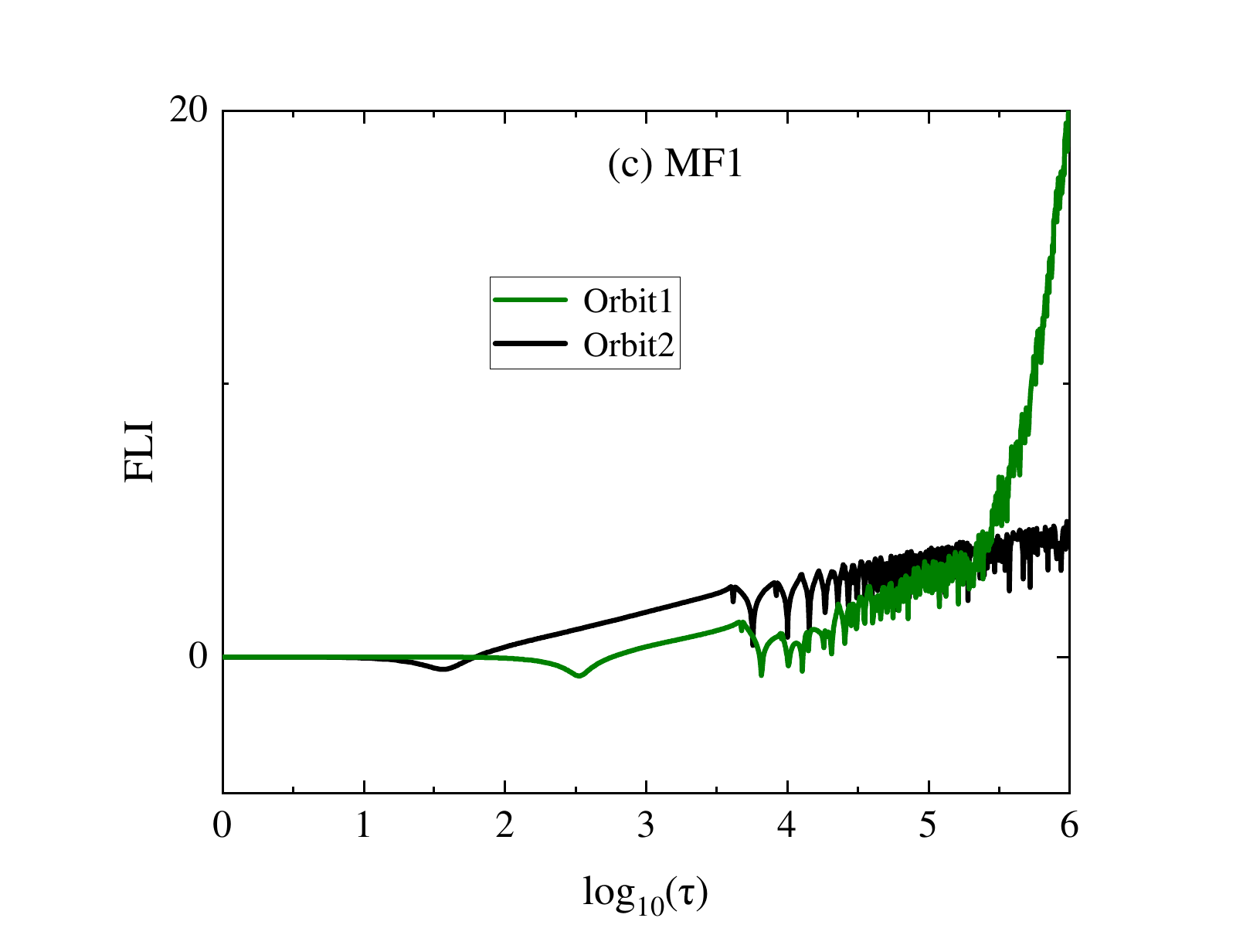}
\includegraphics[width=13pc]{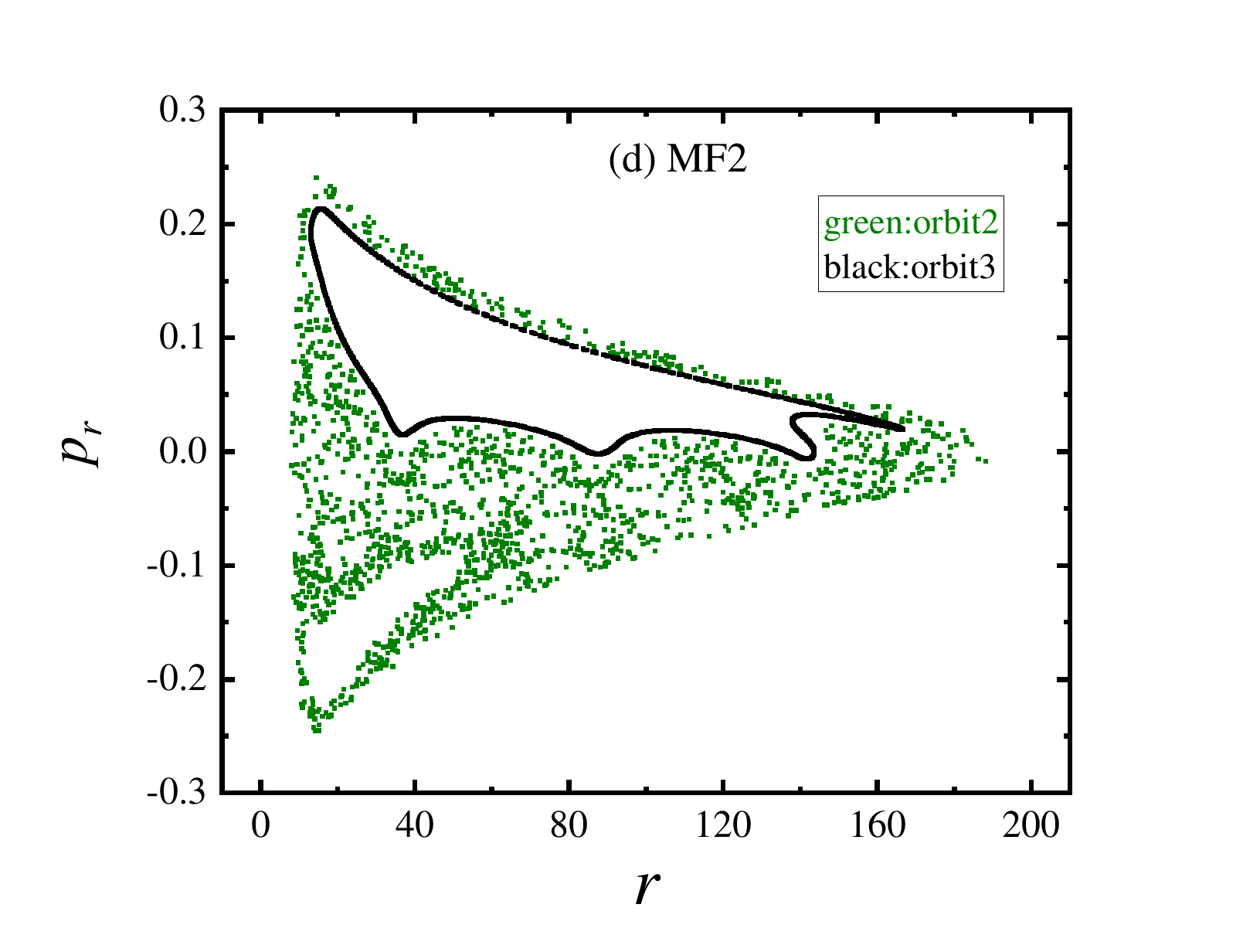}
\includegraphics[width=13pc]{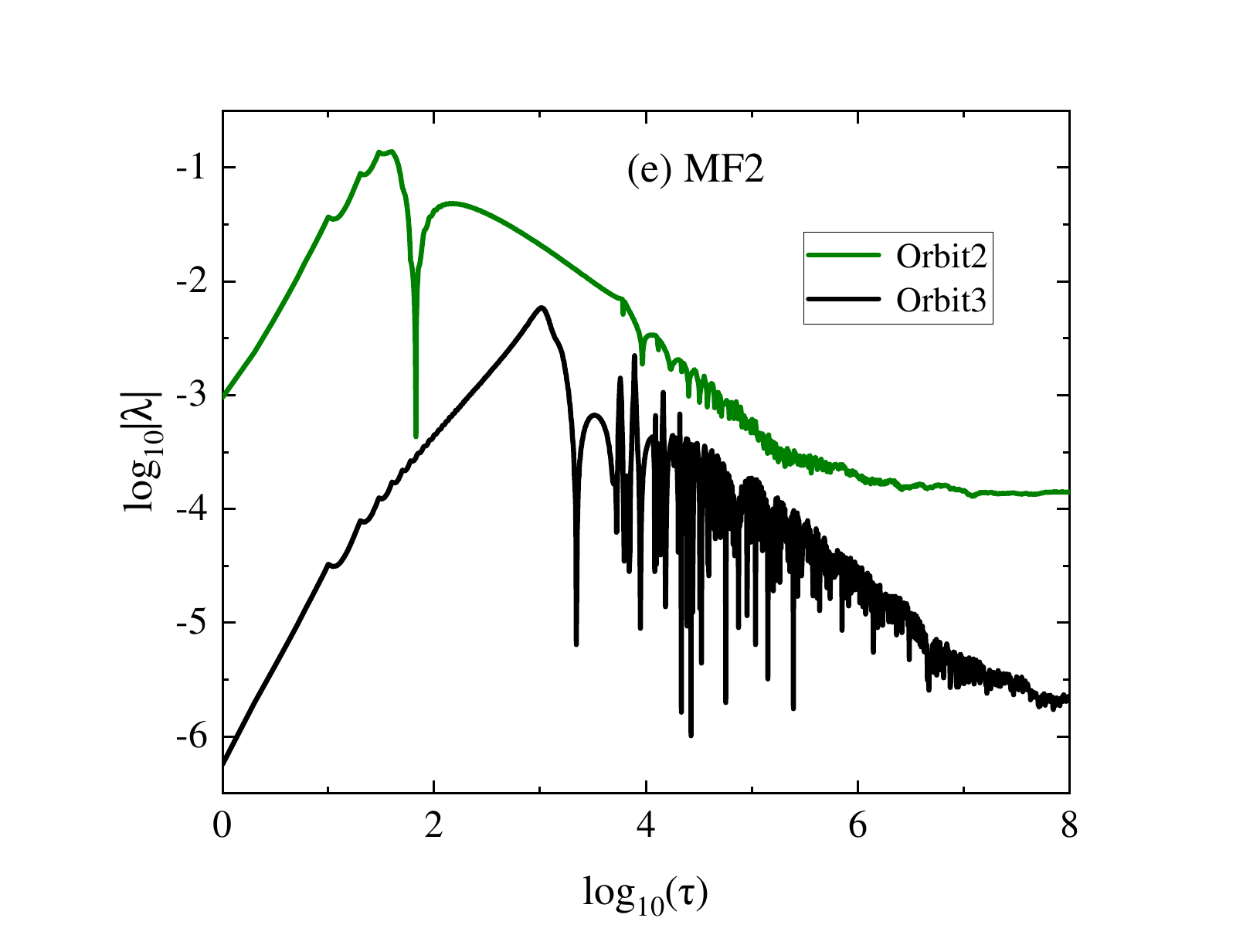}
\includegraphics[width=13pc]{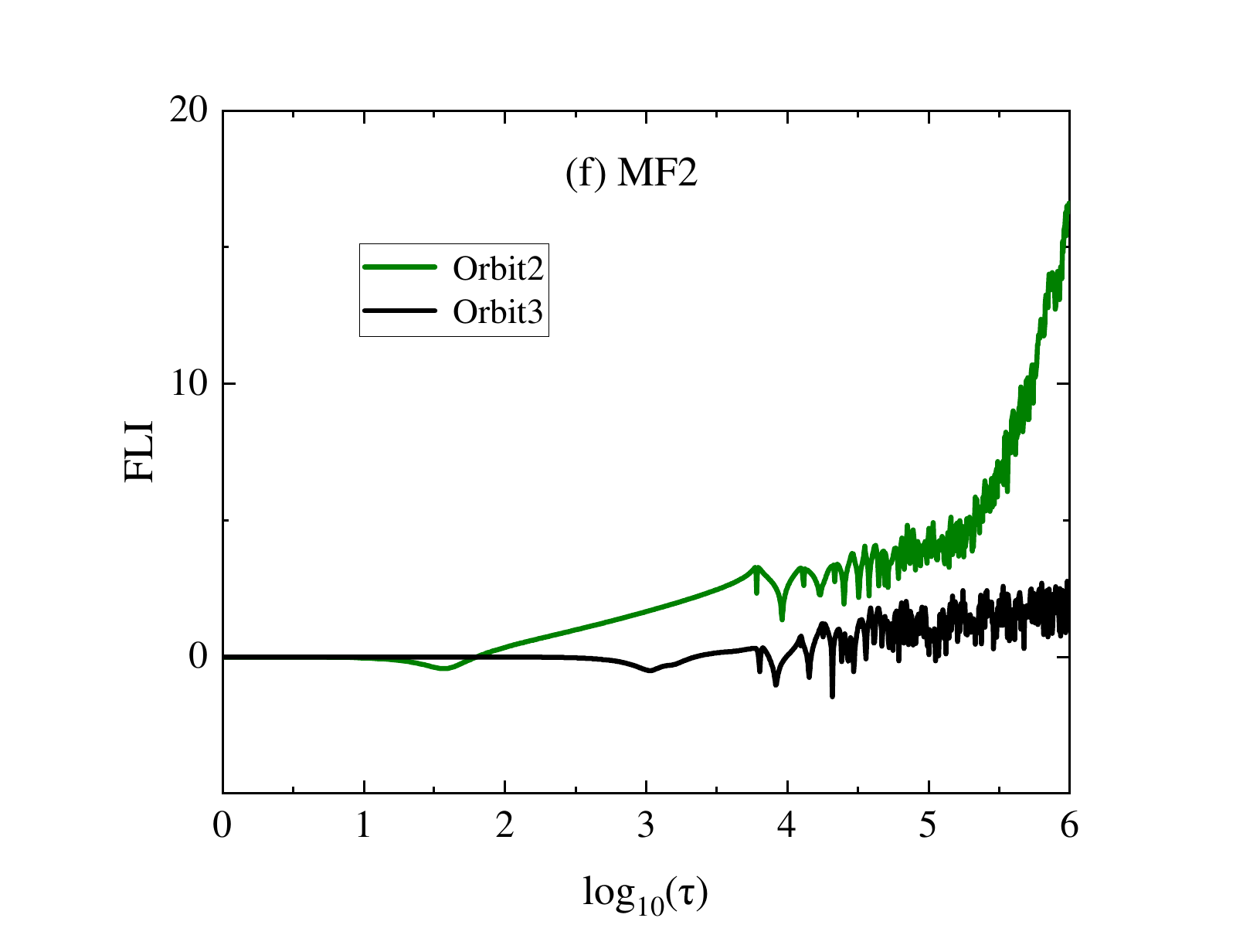}
\caption{(a) Poincar\'{e} sections at the plane $\theta =\pi/2$
with $p_\theta > 0$ for chaotic orbit 1 and regular orbit 2 in MF1
of Figure 1(a),(b). (b) The largest Lyapunov exponents (LEs)
$\lambda$ for the two orbits in panel (a). (c) The fast Lyapunov
indicators (FLIs) for the two orbits in panel (a). (d)
Poincar\'{e} sections for chaotic orbit 2 and regular orbit 3 in
MF2 of Figure 1(c),(d). (e) The LEs $\lambda$ for the two orbits
in panel (d). (f) The FLIs for the two orbits in panel (d).
Clearly, the LEs and the FLIs are the same as the method of
Poincar\'{e} sections in the description of dynamical features of
the three orbits. }}
\end{figure*}

\begin{figure*}[htpb]
\centering{
\includegraphics[width=13pc]{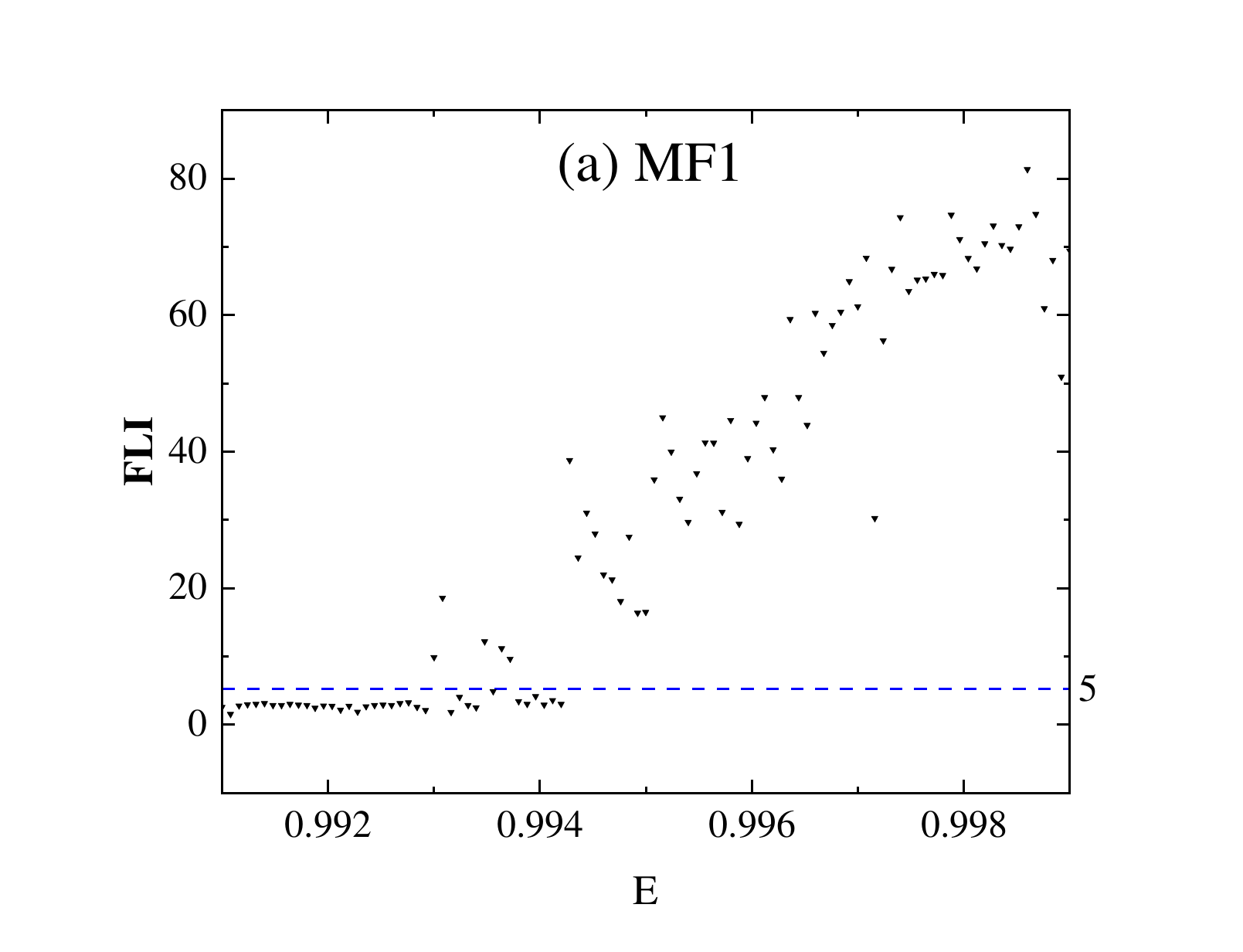}
\includegraphics[width=13pc]{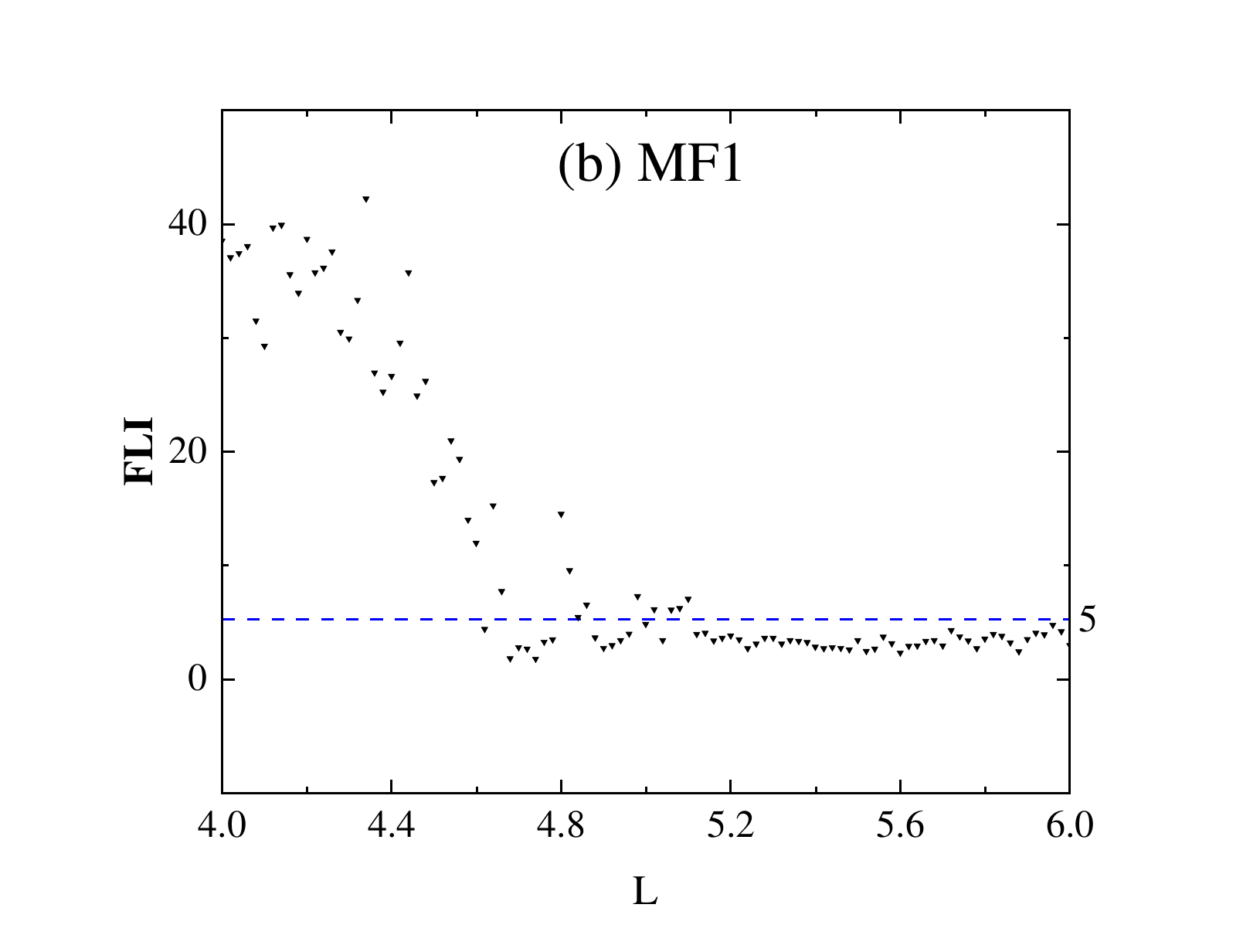}
\includegraphics[width=13pc]{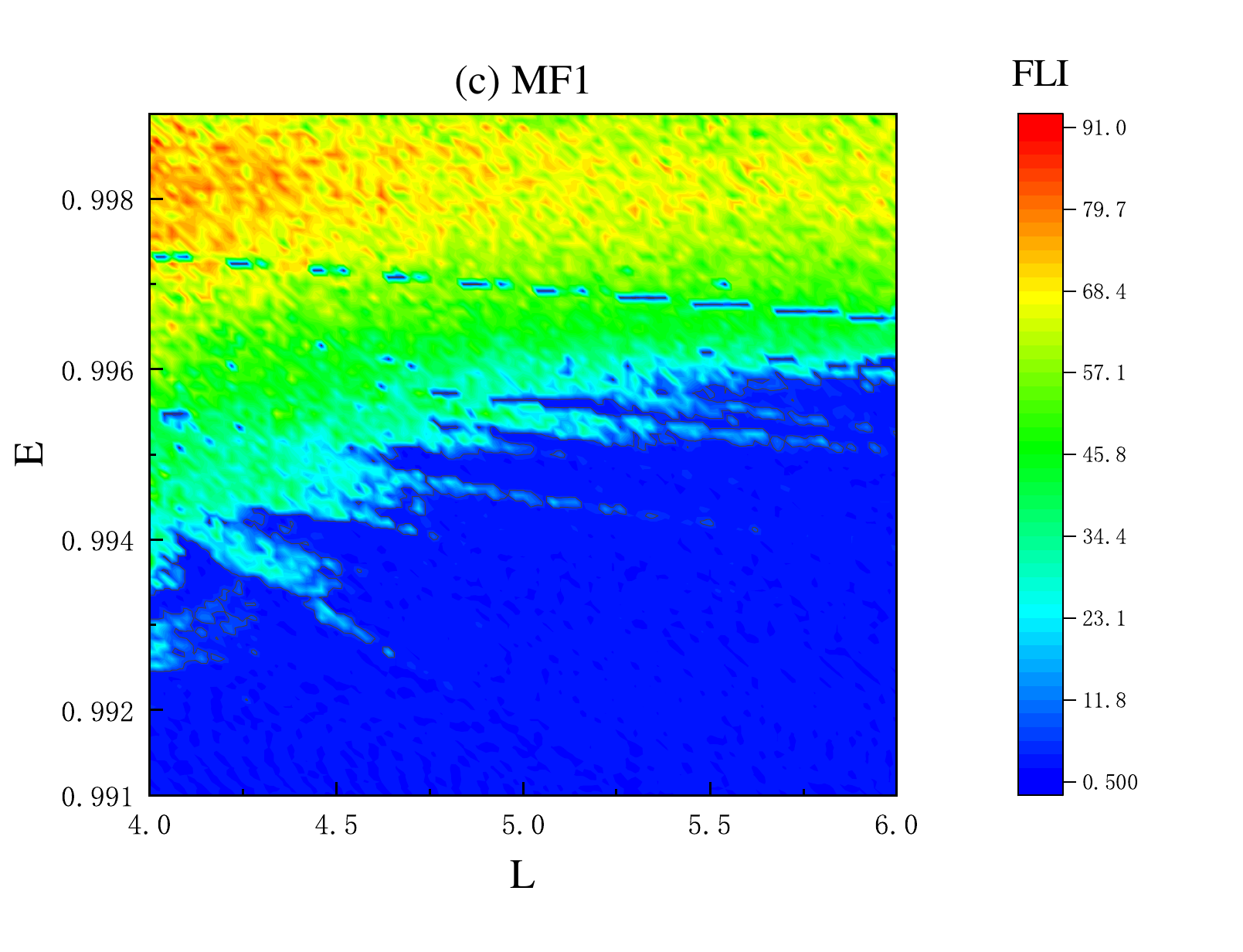}
\includegraphics[width=13pc]{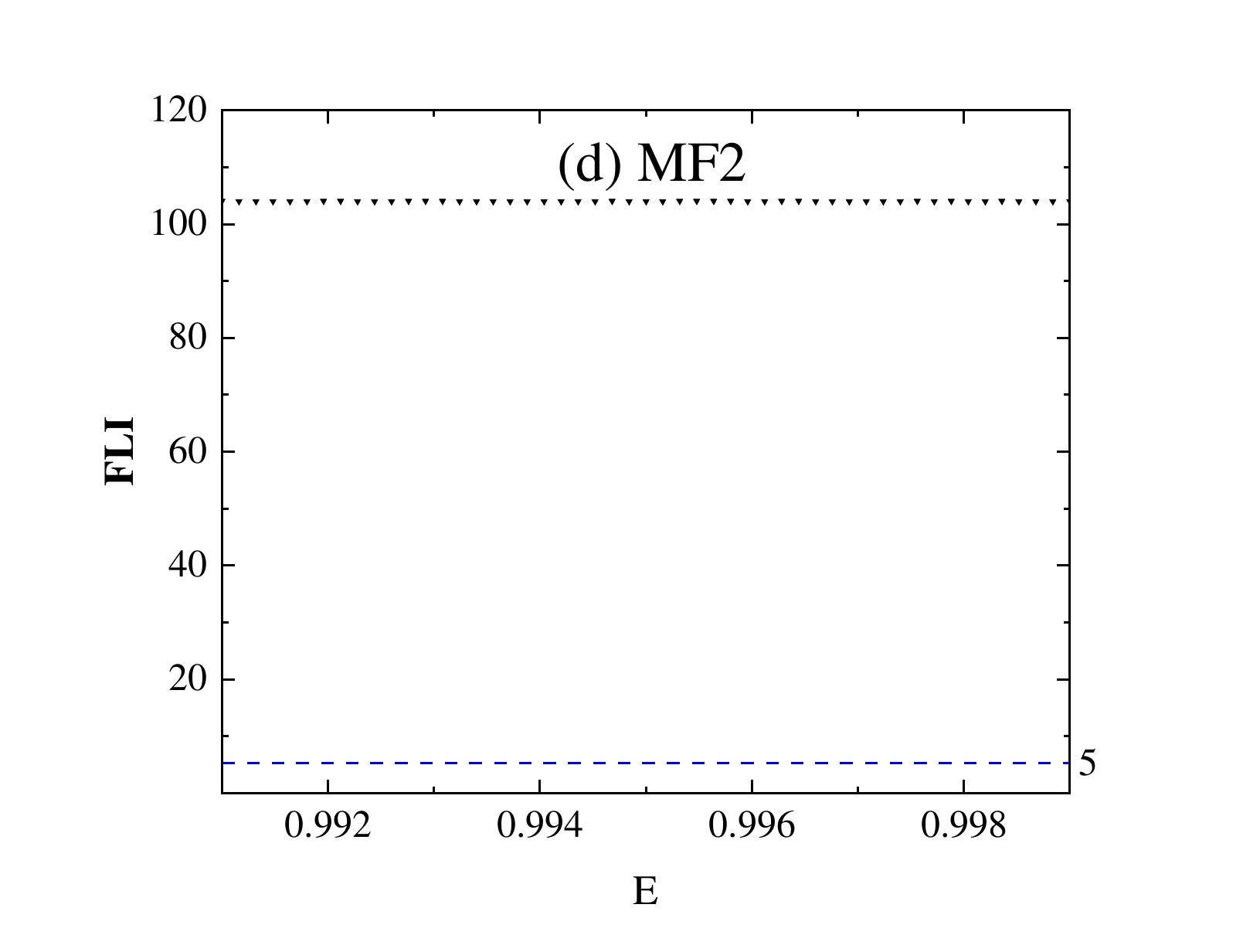}
\includegraphics[width=13pc]{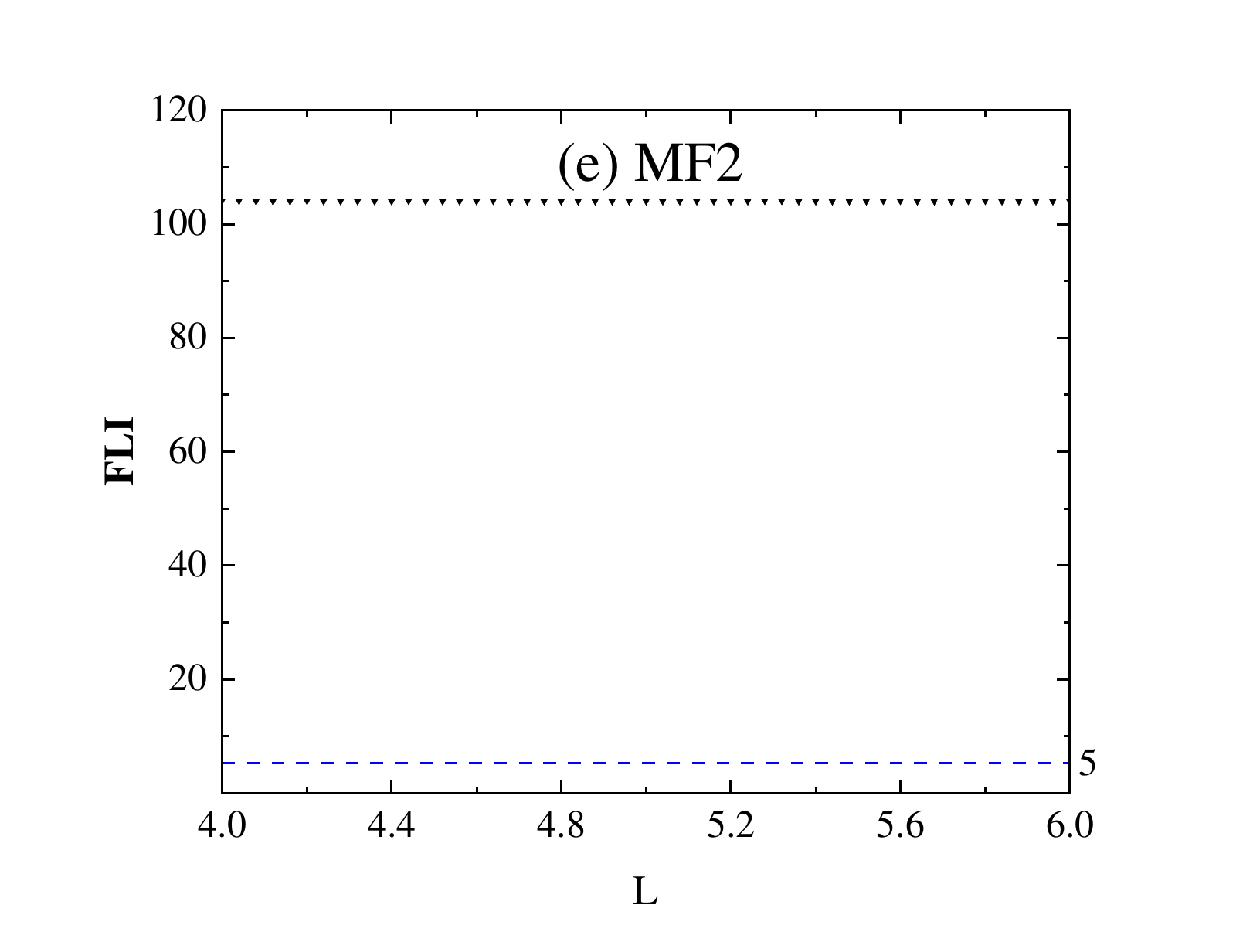}
\includegraphics[width=13pc]{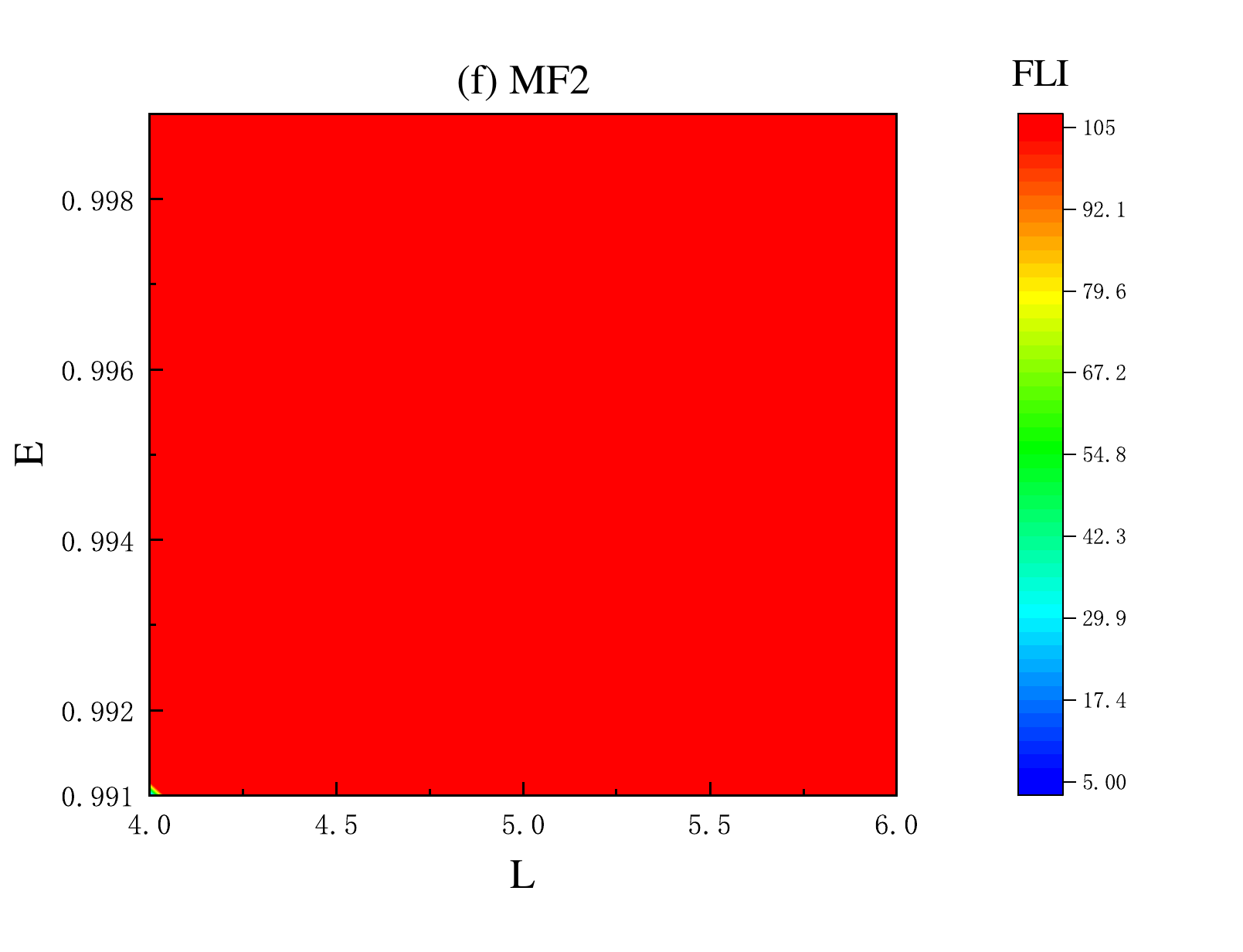}
\caption{Dependence of the FLIs on the parameters $E$ and/or $L$.
The magnetic field parameter is $\beta=1.85\times 10^{-3}$, and
the initial separation is $r =50$. (a): Effect of $E$ on the FLI
with $L = 4.5$ in MF1. Chaos is most likely to happen and is
strengthened as $E>0.993$ increases. (b): Effect of $L$ on the FLI
with $E = 0.995$ in MF2.  Chaos is most likely to occur and to be
strengthened as $L<5.1$ decreases. (c): Finding regular and
chaotic regions in the two parameter space $(E,L)$ using the FLIs
in MF1. The effects of $E$ and $L$ on chaos are consistent with
those in panels (a) and (b). Panels (d)-(f) respectively
correspond to panels (a)-(b), but MF2 is used instead of MF1.
Strong chaos is always allowed in this case.
 } }
\end{figure*}

\begin{figure*}[htpb]
\centering{
\includegraphics[width=16pc]{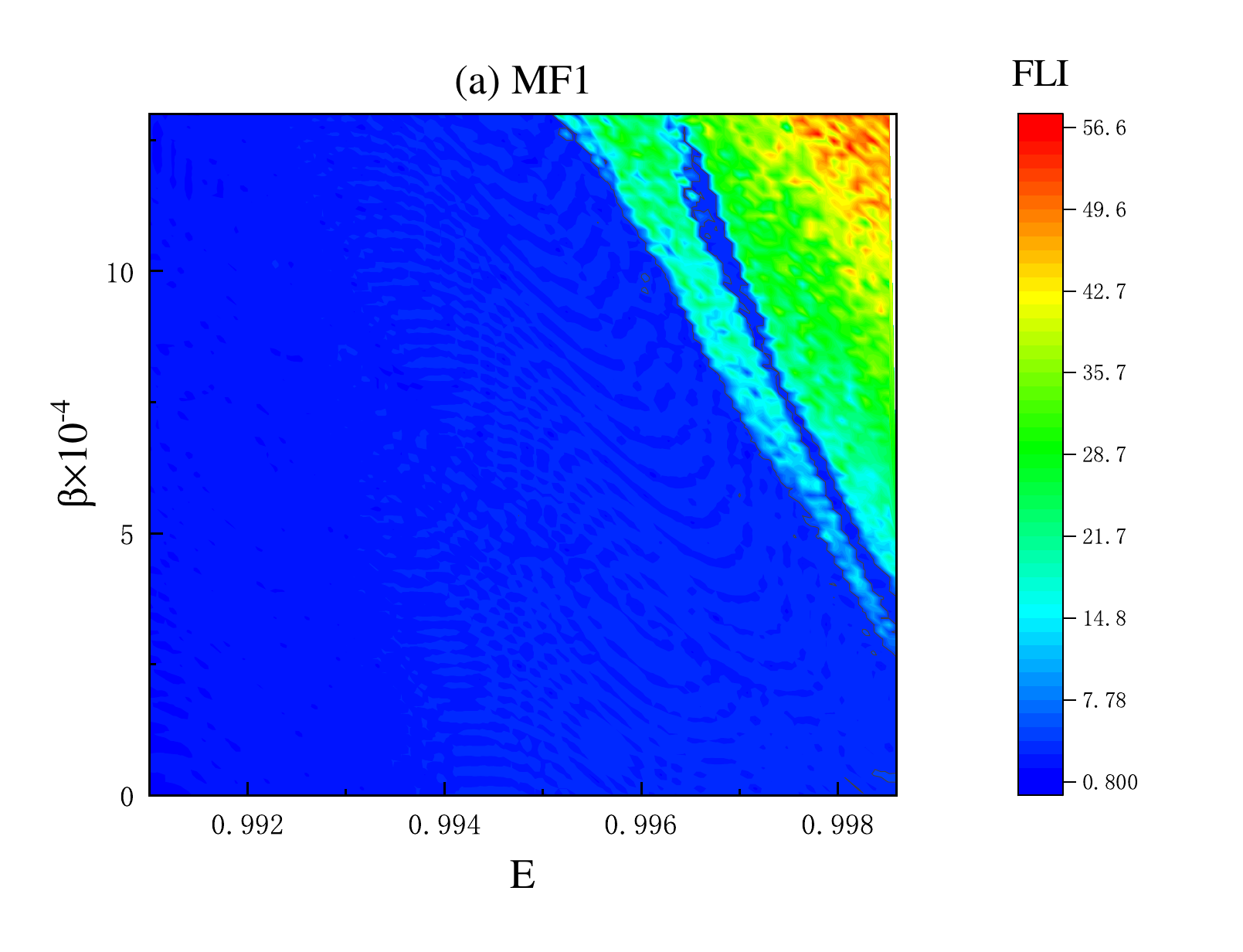}
\includegraphics[width=16pc]{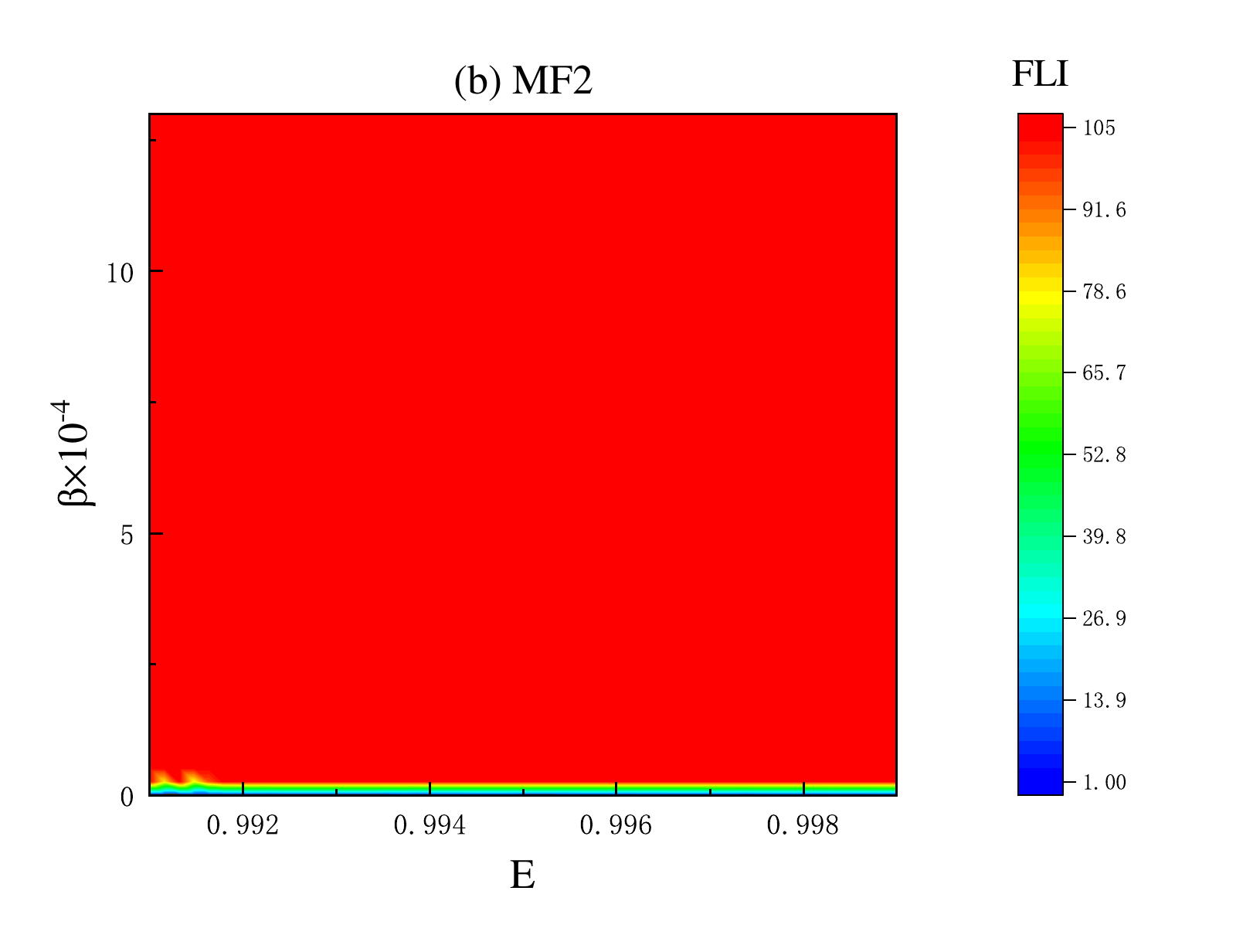}
\caption{The parameters $(E,\beta)$ corresponding to the FLIs,
which indicate regular and chaotic dynamics. The angular momentum
is $L = 4.3$, and the initial separation is $r = 50$. (a): Chaos
gets stronger as both $E$ and $\beta$ increase in MF1. (b): Strong
chaos is almost existent in MF2. } }
\end{figure*}
\begin{figure*}[htpb]
\centering{
\includegraphics[width=16pc]{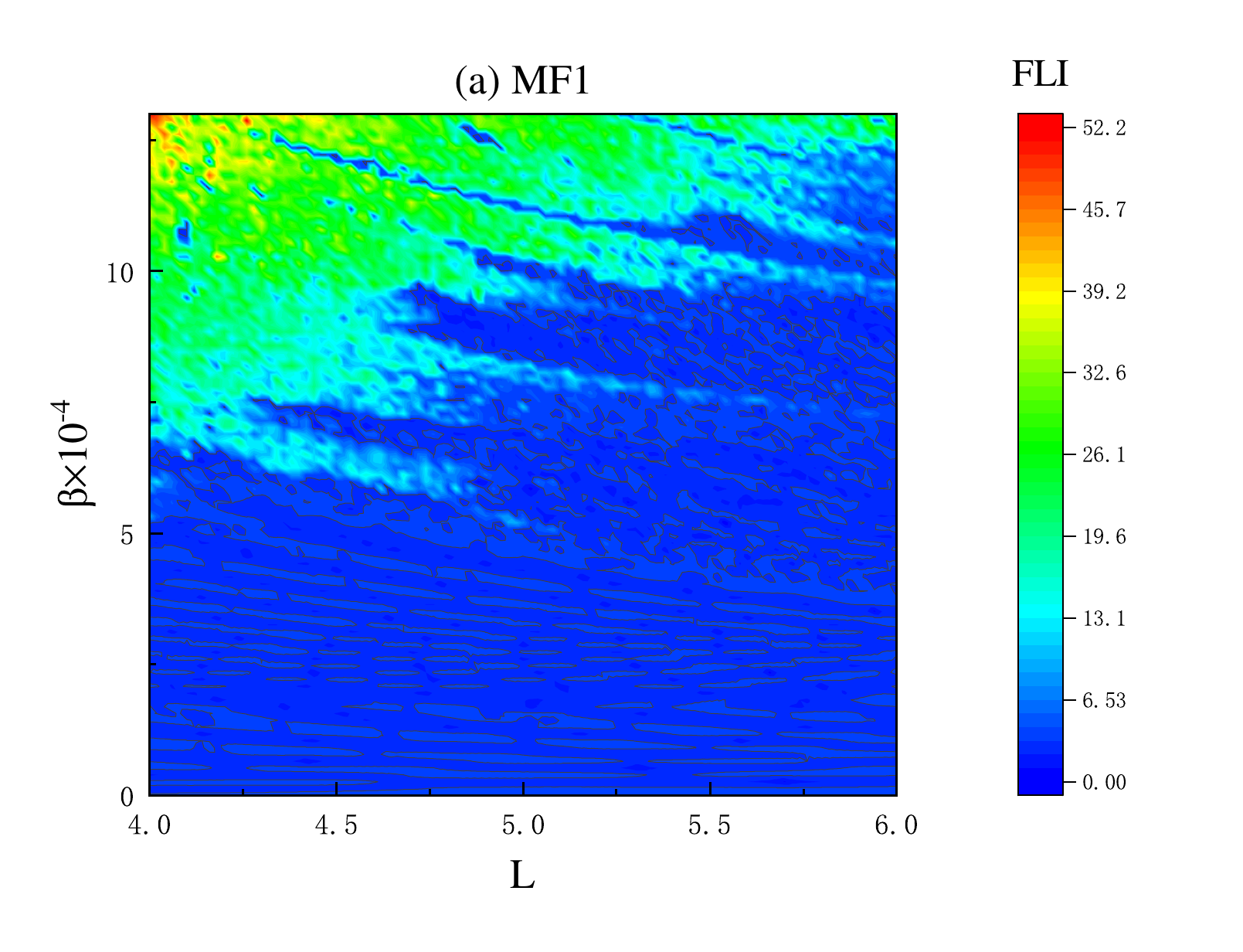}
\includegraphics[width=16pc]{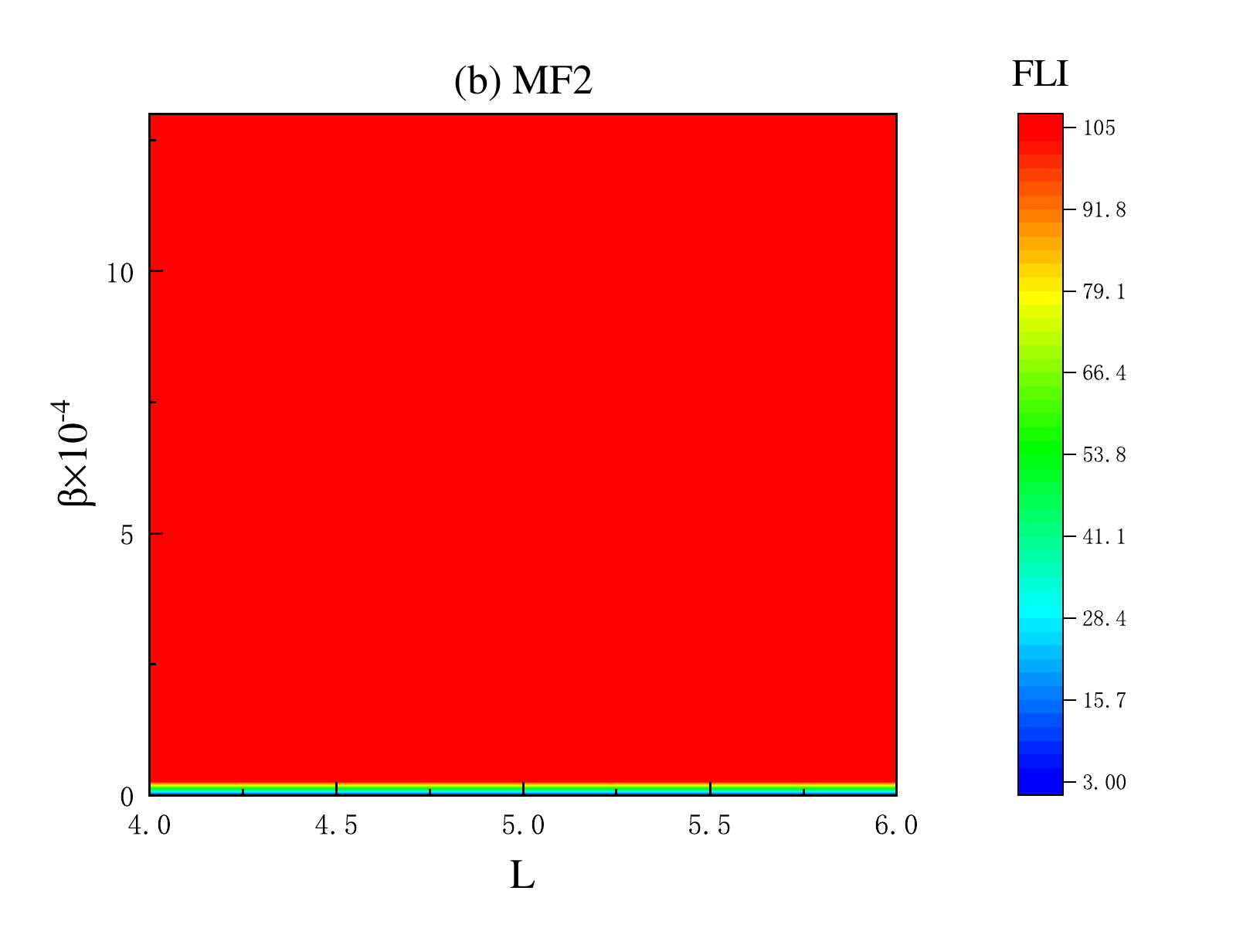}
\caption{The parameters $(L,\beta)$ corresponding to the FLIs,
which indicate regular and chaotic dynamics. The energy is $E =
0.997$, and the initial separation is $r=70$. (a): Chaos gets
stronger as $L$ is smaller and $\beta$ is larger in MF1. (b):
Strong chaos is almost existent in MF2. } }
\end{figure*}

\end{document}